\documentclass[pre]{revtex4}
 \usepackage[english]{babel}
 \usepackage{amsmath}
 \usepackage{epsfig}
\begin{document}

\title{ANALYSIS OF LINEAR AND NONLINEAR CONDUCTIVITY OF PLASMA-LIKE SYSTEMS ON THE BASIS OF THE FOKKER-PLANCK EQUATION}
\author{S.A. Trigger $^{1,2,3}$, W. Ebeling $^2$, G.J.F. van Heijst $^3$, D. Litinski $^3$}
\address{$^1$ Joint\, Institute\, for\, High\, Temperatures, Russian\, Academy\,
of\, Sciences, 13/19, Izhorskaia Str., Moscow\, 125412, Russia; email:\, satron@mail.ru\\
$^2$  Institut f\"ur Physik, Humboldt-Universit\"at zu Berlin,
Newtonstra{\ss}e 15, D-12489 Berlin, Germany\\
$^3$ Eindhoven\, University\, of\, Technology, P.O. Box 513, MB 5600
Eindhoven, The Netherlands\\}

\begin{abstract}

The problems of high linear conductivity in an electric field, as well as nonlinear conductivity, are considered for plasma-like systems.

First, we recall several observations of nonlinear fast charge transport in dusty plasma, molecular chains, lattices, conducting polymers and semiconductor layers. Exploring the role of noise we introduce the generalized Fokker-Planck equation.

Second, one-dimensional models are considered on the basis of the Fokker-Planck equation with active and passive velocity-dependent friction including an external electrical field.
On this basis it is possible to find the linear and nonlinear conductivities for electrons and other charged particles
in a homogeneous external field.
It is shown that the velocity dependence of the friction coefficient can lead to an essential increase of the electron average velocity and the corresponding conductivity in comparison with the usual model of constant friction, which is described by the Drude-type conductivity. Applications including novel forms of controlled charge transfer and non-Ohmic conductance are discussed.\\

PACS number(s): 05.20.Dd, 05.60.Cd, 52.25.Fi, 52.65.Ff

\end{abstract}

\maketitle

\section{Introduction}

According to the classical conductance theory due to Paul Drude, Ohmic currents are
proportional to external field and inversely proportional to the friction constant $m \gamma_0$
[1], where $m$ is the mass and $\gamma_0$ the collision frequency.

The elementary Drude theory calculates the drift velocity $v_D$ from the equilibrium between
electrical field and friction forces:
\begin{equation}
e E = m \gamma_0 v_D; \qquad v_D = \frac{e E}{m \gamma_0},\label{A1}
\end{equation}
where $\gamma_0$ is the velocity-independent friction coefficient.
The Drude current and the corresponding conductivity are, therefore, equal to
\begin{equation}
j_D = n e v_D =  \frac{n e^2 E}{m \gamma_0}, \qquad \sigma_D = \frac{n e^2}{m \gamma_0}.
\label{A2}
\end{equation}
Many systems do not obey such a simple dependence but show complex nonlinear dependence.
For electrolytes a strong increase of the conductivity with the field was first observed by Max Wien and is known as
the Wien effect, and Hans Falkenhagen and Lars Onsager have contributed to the theoretical interpretation, see
[2].
Nonlinear effects in strong electric fields are also known from plasma physics
[3],[4]. Further examples of nonlinear conduction phenomena
were studied experimentally and theoretically for special polymers [5],[6].
In dusty plasmas nonlinear effects can be observed, in particular, in relation to the ion drag force [7].

In typical Drude-like conductors as electrolytes, partially ionized plasmas,
semiconductors and metals charge velocities at a typical field strength of
1 V/cm are smaller than 1 m/s. In the more exotic conductors we have in mind here as
special PDA-polymers [5] and in dusty plasmas the characteristic velocities can reach
up to 1000 times higher. For this reason the field values we explore may be potentially important for the
development of new fast conductors very different from superconductors.

Typically high drift velocities are nonlinear and non-Ohmic.
Non-Ohmic high drift velocities depending on the field strength were observed experimentally in
many different systems [5],[7].

Theoretical models were developed in part based on Fokker-Planck models with complicated friction and diffusion functions including negative friction [8]-[10].
Nonlinear non-Ohmic conductance phenomena were studied experimentally and theoretically also in nonlinear driven electric circuits [11]-[13].
Different approaches in the framework of polaron theory were given in
[14]-[19], and this list could easily be extended.

We note that the low-field drift corresponds to a high conductivity which may be much higher then the
Drude conductivity.
Here we study the conductivity in the framework of Fokker-Planck theory.

\section{Fokker-Planck equation in a homogeneous external electric field}

In the general homogeneous case for collisions, leading to normal diffusion, we use the equation for particles with charge -$e$ (e is positive) and mass $m$:
\begin{equation}
\frac{\partial f({\bf v},t)}{\partial t}-\frac{e{\bf E}}{m}\frac{\partial f({\bf v},t)}{\partial {\bf v}} = \frac{\partial}{\partial {\bf v}_i}\left[\gamma(v){\bf v}_i f({\bf v},t)+ \frac{\partial}{\partial {\bf v}_j} D_{i,j}({\bf v}) f({\bf v},t)\right].
\label{A3}
\end{equation}
Let us consider pure formally the one-dimensional case. The distribution is the sum of two parts: $f^+$ with velocity parallel to $E$, and $f^-$ with velocity antiparallel to $E$:
\begin{equation}
\frac{\partial f^\alpha(v,t)}{\partial t}-\alpha\frac{eE}{m}\frac{\partial f^\alpha(v,t)}{\partial v} = \frac{\partial}{\partial v} \left[\gamma(v) v f^\alpha(v,t)+ \frac{\partial}{\partial v} D( v) f^\alpha (v,t)\right],
\label{A4}
\end{equation}
where $\alpha=\pm$ and everywhere the variable $v$ changes from $0$ to $\infty$, since the direction of velocity is taken into account by the sign of $\alpha$.

For $E=0$ the stationary solution $f_s(v)$ is not dependent on $\alpha$ ($f^+_s=f_s^-=f^0_s(v)$) and reads:
\begin{eqnarray}
f^\alpha_s (v,t)\equiv f^0_s(v)=\frac{C}{D(v)}\exp \left(-\int \frac{v \gamma(v)}{D(v)}dv \right).
\label{A5}
\end{eqnarray}
Here $C$ is the normalization constant.

Since the values $\gamma(v)$ and $D(v)$ depend on modulus $v$ and $E=0$ we arrive at the normalization
condition
\begin{eqnarray}
\int_0^{\infty} f^-_s(v)dv+\int_0^{\infty} f^+_s(v)dv=2\int_{0}^\infty f^0_s(v)dv=n; \nonumber\\
n=2 C \int_0^\infty  \frac{dv}{D(v)}\exp \left(-\int_0^v \frac{v' \gamma(v')}{D(v')}dv' \right).
\label{A6}
\end{eqnarray}

If $\gamma(v)$ and $D(v)$ are not dependent of $v$ we find that the equilibrium condition satisfies the
Einstein relation $\gamma_0 /D_0=m/T$ and normalization leads to the following equality ($x\equiv v^2/v_0^2$, $v^2_0\equiv 2D_0/\gamma_0$):
\begin{equation}
C=n D_0 \sqrt{\frac{m}{2\pi T}} \qquad f^0_s=n  \sqrt{\frac{m}{2\pi T}} \exp \left(-  \frac{ \gamma_0 v^2 }{2 D_0}\right)= \frac{n}{\sqrt \pi v_0 } \exp \left(-x^2 \right) .
\label{A7}
\end{equation}

\section{Influence of a homogeneous external electric field: general solution in one-dimensional case}

Let us generalize solution (5) for the case of the system in an external homogeneous field. The general stationary
solution can be easily found for arbitrary $v$-dependence of the friction and diffusion coefficients $\gamma(v)$ and $D(v)$  by substitution the value $v\gamma(v)+\alpha eE/m$ instead $v\gamma(v)$
\begin{equation}
f^\alpha_s(v)= \frac{C'^\alpha}{D(v)}\exp \left(-\int \frac{v \gamma(v)+\alpha eE/m}{D(v)}dv \right).
\label{A8}
\end{equation}
For the case of constant $\gamma(v)=\gamma_0$ and $D(v)=D_0$ (this case corresponds to equilibrium if the
Einstein relation $\gamma_0 /D_0=m /T$ is fulfilled and the electric field $E=0$)  the distribution function is equal to
\begin{equation}
f^\alpha_s(v)= \frac{C'^\alpha}{D_0}\exp \left(- \frac{v^2 \gamma_0}{2D_0}-\alpha\frac{eE v}{m D_0}\right).
\label{A9}
\end{equation}

By using the normalization condition we find
\begin{equation}
C'^+=C'^-\equiv C'\, , \qquad C'=\frac{n \sqrt{D_0 \gamma_0}}{\sqrt{2 \pi}}
exp(-e^2E^2/2 \gamma_0 m^2 D_0).
\label{A10}
\end{equation}

For the case of a weak electric field (7) reads as
\begin{equation}
f^\alpha_s(v)= \frac{C'}{D_0}\exp \left(- \frac{v^2 \gamma_0}{2D_0}\right)[1-\alpha\frac{eE v}{m D_0}].
\label{A11}
\end{equation}
where $C'=C'(E=0)$ due to linearity of the approximation is determined by equality
\begin{equation}
C'(E=0)=\frac{n \sqrt{D_0 \gamma_0}}{\sqrt{2 \pi}},
\label{A12}
\end{equation}
which we have to use for calculating of the current.

Then, the current in the linear approximation (11),(12) equals
\begin{eqnarray}
j=-e \sum_\alpha \int dv v \alpha f^\alpha_s(v)= -e \int dv v [ f^+_s(v)-f^-_s(v)]=\nonumber\\e^2 E\frac{2n \sqrt{ D_0\gamma_0/2\pi}}{m D^2_0}\int_0^\infty dv v^2 \exp \left(- \frac{v^2 \gamma_0}{2D_0}\right)=E e^2n/m \gamma_0.
\label{A13}
\end{eqnarray}
This result corresponds to the Drude formulation.

Now we calculate the nonlinear stationary current
\begin{eqnarray}
j=
-e \frac{C'}{D_0}\int_{-\infty}^\infty dv v \exp \left(- \frac{v^2 \gamma_0}{2D_0}-\frac{e E v}{m D_0}\right)=
e^2 E \frac{C'}{ m  \gamma_0}\sqrt{\frac{2 \pi}{\gamma_0 D_0}}\exp [\frac{e^2 E^2 }{2 m^2 D_0\gamma_0}].
\label{A14}
\end{eqnarray}
By use of the normalization function (10) we again arrive at the Drude result
\begin{equation}
j= e^2 n E/m \gamma_0.
\label{A15}
\end{equation}

Therefore, the Fokker-Planck equation for the one-dimensional case with constant coefficients $\gamma_0$ and $D_0$ even for a strong homogeneous electric field leads to a current that is linearly dependent on $E$ and to the Drude conductivity, although the velocity distribution is a nonlinear function of $E$.

On other hand, we know that in a plasma there are running electrons due to a decrease of the friction force at high velocities (which behaves as $1/v^2$, see, e.g., [20]). Therefore, to describe real systems we have to extend our consideration on the case of the velocity-dependent friction coefficient. Such type of extension can be applied to various physical systems, as, e.g., plasmas, or polarons in solid matter, or dusty particles in a dusty plasma. For each case we should specify the particular velocity dependence of the friction coefficient. An example of such a system is considered in the next section.

\section{The model with a velocity-dependent passive friction}

On the basis of the previous arguments in the classical case we have consider a non-equilibrium situation to deviate from the Einstein relation and from the picture described above.

Let us consider the model of friction for the classical non-equilibrium stationary system of charged particles, when the friction coefficient is positive for all velocities (so-called passive friction)
\begin{equation}
\gamma(v)=\gamma_0\frac{1+\mu v^2}{1+\beta v^4}, \qquad D=D_0.
\label{A16}
\end{equation}
Then the distribution (5) (for the case $E=0$) reads
\begin{equation}
f_s(v)=\frac{C}{D_0}\frac{1}{(1+\beta v^4)^{\gamma_0 \mu /4\beta D_0}}\exp \left(-\frac{\gamma_0}{2D_0\sqrt \beta}arctg \sqrt{\beta}\,v^2 \right),
\label{A17}
\end{equation}
where $C$ is determined by the normalization condition (the substitution $\zeta\equiv \sqrt{\beta}\,v^2$ is used)
\begin{equation}
C=\frac { n D_0\beta^{1/4}}{\int_0^\infty  \frac{d \zeta }{\zeta^{1/2}(1+\zeta^2)^{\gamma_0 \mu/4\beta D_0}}\exp \left(-\frac{\gamma_0}{2D_0\sqrt \beta}arctg \zeta \right)}.
\label{A18}
\end{equation}
The limiting forms for the function $f_s(v)$ are (it is easy to verify that the limits $\beta\rightarrow 0$ and $v\rightarrow\infty$ are not transposed)
\begin{equation}
\lim_{v\rightarrow \infty} f_s(v)=\frac{C}{D_0}\frac{1}{(1+\beta v^4)^{\gamma_0 \mu/4\beta D_0}}\exp \left(-\frac{\gamma_0\sqrt{\pi}}{2{\sqrt {2\beta}} D_0} \right)\simeq \frac{C'}{v^{\gamma_0 \mu/\beta D_0}},
\label{A19}
\end{equation}
\begin{equation}
\lim_{v\rightarrow 0} f_s(v)\simeq\frac{C}{D_0}\exp \left(-\frac{\gamma_0 v^2}{2D_0} \right)\simeq\frac{C}{D_0}\left(1-\frac{\gamma_0 v^2}{2 D_0 } \right).
\label{A20}
\end{equation}
Obviously the distribution has a long tail in velocity space and in this sense is anomalous ([21]-[24]).

\begin{figure}
\begin{center}
\includegraphics[width=6cm,height=6cm]{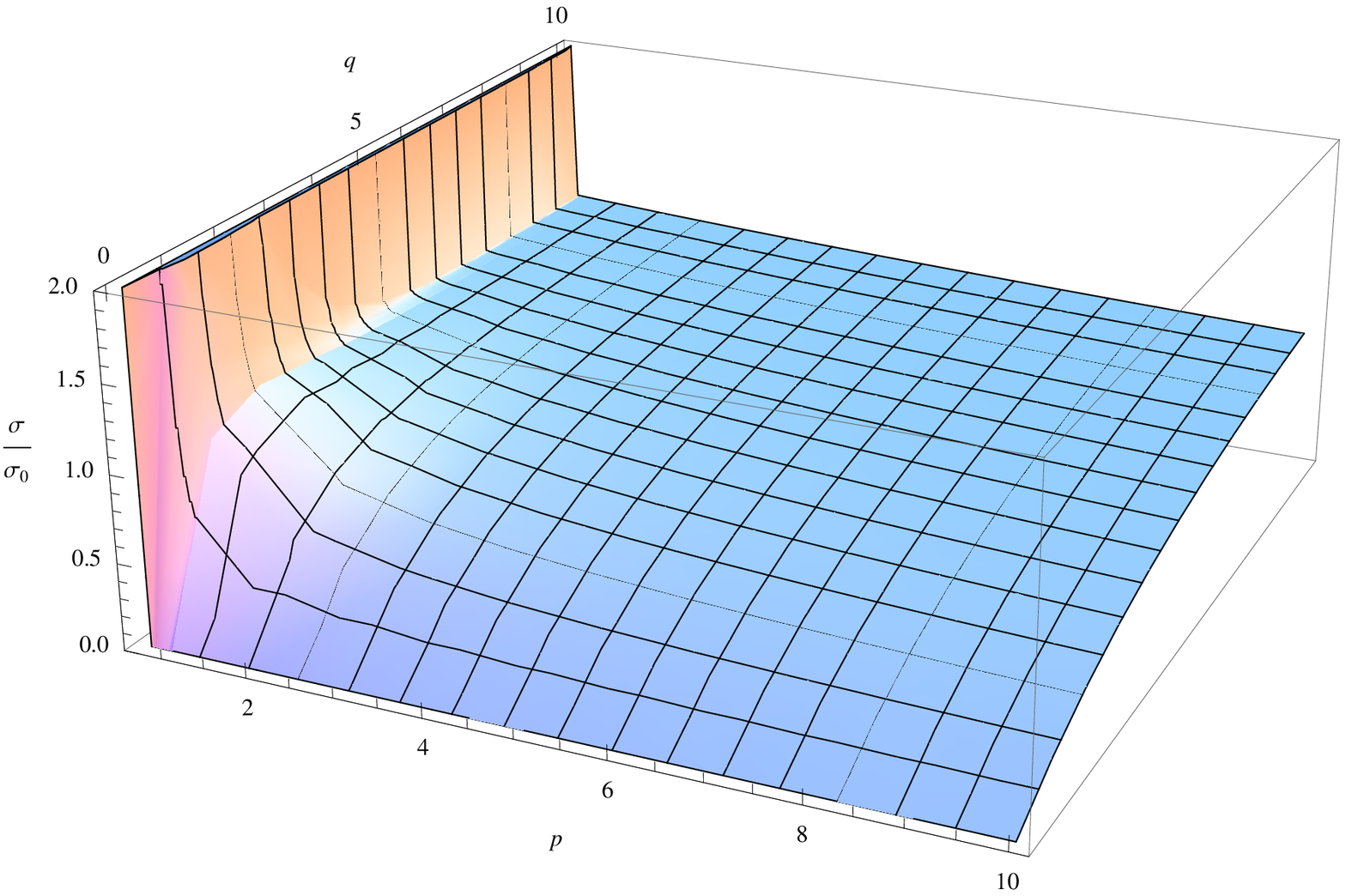}
\includegraphics[width=6cm,height=6cm]{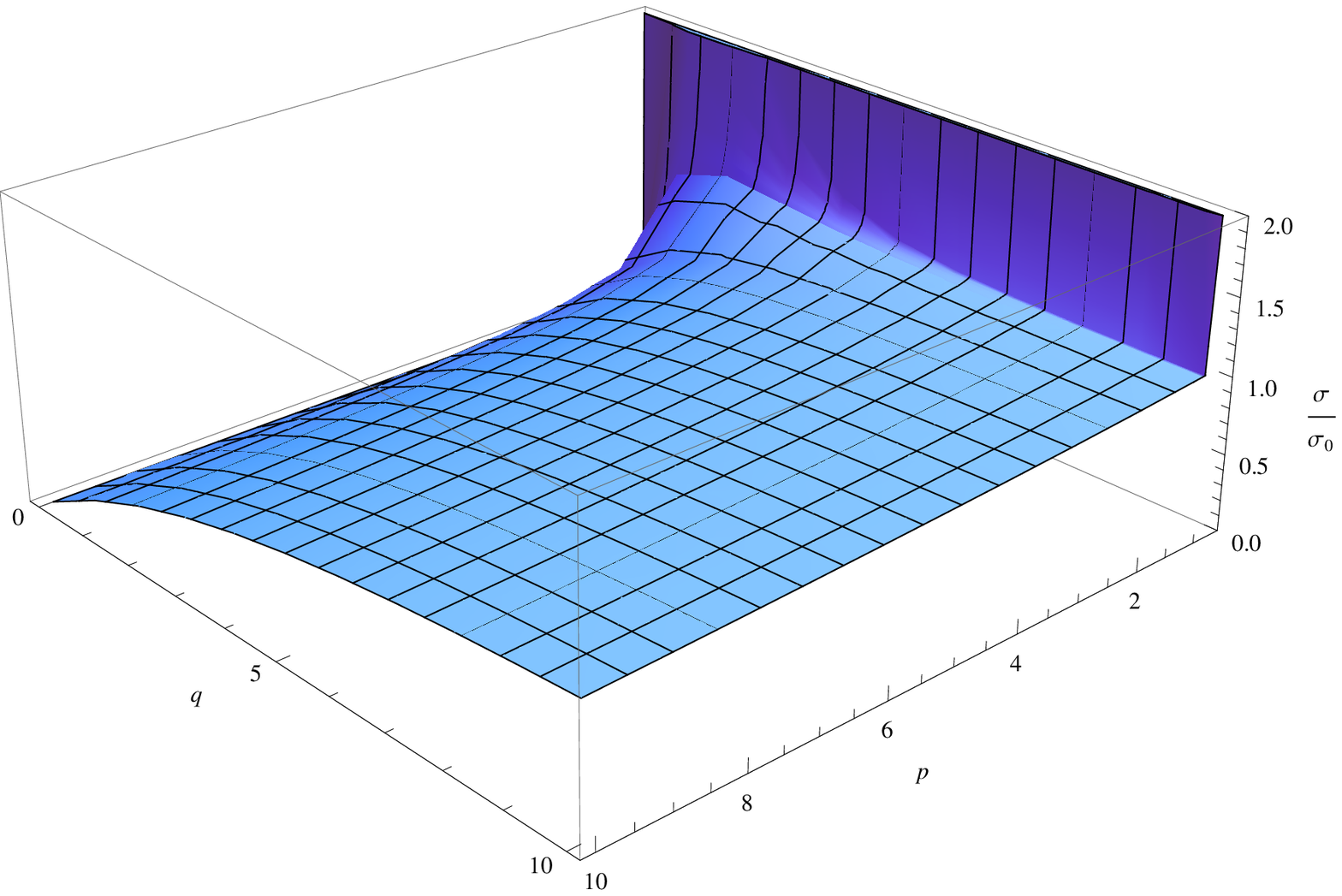}
\end{center}
\caption{Graphical representation of the dependence $\sigma/\sigma_D$ as a function of the parameters $p$ and $q$, plotted for two different viewing angles.}
\label{Fig.1}
\end{figure}






To find the stationary solution in a homogeneous electric field we have to apply the substitution $\gamma (v)\rightarrow \gamma(v)+(\alpha eE/m v)$ in (3) and then calculate the distribution. Instead (9), taking into account (8) (for $D=D_0$), we find the general (nonlinear on $E$) solution for the considered form (16) of the friction $\gamma(v)$:
\begin{equation}
f_s^\alpha(v)=\frac{C^\alpha}{D_0}\frac{1}{(1+\beta v^4)^{\gamma_0 \mu/4\beta D_0}}\exp \left(-\frac{\gamma_0}{2D_0\sqrt \beta}arctg \sqrt{\beta}\,v^2 \right)\times \exp \left(-\frac{\alpha e E v}{m D_0}\right).
\label{A21}
\end{equation}

Let us at first investigate the linear response on the external field $E$.
Then the part of the
distribution $f^{(1)}_s(v)$, which determines the linear conductivity reads
\begin{equation}
f^{(1\alpha)}_s(v)=- \alpha E \frac{e C}{m D^2_0}\frac{v}{(1+\beta v^4)^{\gamma_0 \mu/4\beta D_0}}\exp \left(-\frac{\gamma_0}{2D_0\sqrt \beta}arctg \sqrt{\beta}\,v^2 \right),
\label{A22}
\end{equation}
and the current equals
\begin{eqnarray}
j=E e^2  \frac{C}{m D^2_0 \beta^{3/4}}\int_{0}^\infty d\eta \frac{\eta^{1/2}}{(1+\eta^2)^{\gamma_0 \mu/4\beta D_0}}\exp \left(-\frac{\gamma_0}{2D_0\sqrt \beta} arctg\,\eta\right),
\label{A23}
\end{eqnarray}
where the normalization constant $C^\alpha=C$ is determined by Eq. (18).

The conductivity $\sigma\equiv\sigma(p;q)$ for this system can be written in the form
\begin{eqnarray}
\sigma(p;q)= \sigma_D \frac{ \gamma_0 \int_{0}^\infty d\eta \frac{\eta^{1/2}}{(1+\eta^2)^{\gamma_0 \mu/4\beta D_0}}\exp \left(-\frac{\gamma_0}{2D_0\sqrt \beta} arctg\,\eta\right)}{ D_0 \beta^{1/2} \int_0^\infty  \frac{d \zeta }{\zeta^{1/2}(1+\zeta^2)^{\gamma_0 \mu/4\beta D_0}}\exp \left(-\frac{\gamma_0}{2D_0\sqrt \beta} arctg\, \zeta \right)}.
\label{A24}
\end{eqnarray}
If we introduce the dimensionless constants $p=\gamma_0 \mu/4\beta D_0$ and $q=\frac{\gamma_0}{2 D_0\sqrt \beta}$ we find
\begin{eqnarray}
\frac{\sigma(p;q)}{\sigma_D}=  2 q I (p;q); \qquad I(p;q)\equiv \frac{\int_{0}^\infty d\eta \frac{\eta^{1/2}}{(1+\eta^2)^p}\exp \left(-q \; arctg\,\eta\right)}{\int_0^\infty  \frac{d \zeta }{\zeta^{1/2}(1+\zeta^2)^p}\exp \left(-q arctg\, \zeta \right)}.
\label{A25}
\end{eqnarray}
As is easily verify by taking $\beta\rightarrow 0$ ($q\rightarrow\infty$) in the function $I$ only a very small $\zeta, \eta$ are essential and in the exponential function in (25) we have to expand $arctg\,\zeta$ and $arctg\, \eta$. The expressions $(1+\zeta^2)^p$, $(1+\eta^2)^p$ can be replaced by $1$, taking into account that in $p$ the value $\mu/\beta$ is finite if we suppose $\mu=Const \cdot\beta$ (or $\mu/\beta \rightarrow 0$). Then, we arrive at $I(p,q)=1/2q$. This means that such type approximation leads to $\sigma(p;q)=\sigma_D$, as above in Eq. (15). However, this approximation is not relevant, since large velocity values always play a role.

Figure 1 show (in different views) the surface $\sigma(p;q)/\sigma_D=2 q I(p;q)$ (on the figures $\sigma_0\equiv\sigma_D$).
We should mention that for small values of $\zeta, \eta$ the integral contains a singularity in the nominator if $p\leq 0.75$. As can be seen in the figures for large $p$ and $q$ the ratio $\sigma/\sigma_D$ tends to unity.

The above consideration leads to the essential conclusion that the conductivity linearized in
$E$ can increase for certain parameter values of the velocity dependent passive friction (in
comparison with a velocity-independent friction coefficient).

\section{Fokker-Planck theory of driven charges}

\subsection{Stationary distributions of driven particles}

In this section the case of driven particles is considered. For typical driven non-equilibrium systems the friction may be negative at small velocities. Using the Gaussian white noise as fluctuation source, the
distribution function of driven charged particles $f({\bf{r},\bf{v}},t)$ obeys the
Fokker-Planck equation (3),(4) with a velocity-dependent friction coefficient which is negative for some values of the velocity (see, e.g., Figure 2). For typical examples the distribution function deviates from Boltzmann and has
two maxima at some finite velocities $\pm V_0$, as shown in figure 3 (see, e.g., [10],[25],[26]). We assume below a purely thermal noise $D(v) = D_0 \simeq \gamma_0 T/m$ and define a characteristic velocity by the condition
\begin{equation}
\frac{\gamma(V_0)}{D(V_0)} = 0.
\end{equation}
In this case $f_s(v)$ has maxima different from zero.

Several authors as, see, e.g., [27] have proposed to introduce a so-called non-equilibrium potential
sometimes also referred to as stochastic potential $\phi_0(v)$, defined as
\begin{equation}
f_s(v) = C \exp \left( - \phi_0(v) \right),
\label{f1}
\end{equation}
where $C$ is the normalization constant.
In the context of polaron theory, Gogolin introduced the notation effective energy by $\epsilon(v)= T \phi_0(v)$ [6].
Note, that for the quantum-statistical polaron systems in organic chains (see, e.g., [6]), the effective energy spectra
with some minimum different from $v = 0$ was found.
Distribution functions with two symmetrical maxima were also found for certain solectron systems.
In these systems electrons are driven to velocities near to the sound velocity
[11]-[15], [28]. The stochastic potentials (effective energies)
have for these velocities minima different from zero (right panel of Figure 3). It is easily to verified that the stochastic potential $\phi_0(v)$ as well as the effective energies  $\epsilon(v)$ are straightforwardly connected with the functions $\gamma(v)$ and $D(v)$.

\begin{figure}[h]
\centering\includegraphics[width=6cm]{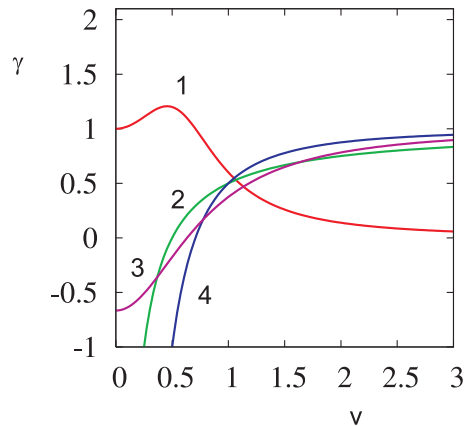}
\caption{{\protect\footnotesize {Graphical representation of different friction functions $\gamma(v)$. Curve 1 corresponds with eq. (16), showing a clear local maximum. The other curves include the value $\gamma (v_i) = 0$: curve 2 represents the Gruler function (29), and curve 3 the SET function given by (31), which corresponds to eq. (30) with $a = 0$. Curve 4 represents the extended SET-model given by eq.(30) with $a>0$.       }}}\label{Fig.2}
\end{figure}


Here we consider only classical models, expecting, however, that the tools developed
here may be applied to a larger spectrum of systems with bistable distribution functions.
In order to proceed we have to specify the functions $\gamma(v), D(v)$.
Several simple formulae for the friction functions of driven particles
are in use. The so called SET-"ansatz" reads [26]
\begin{equation}
\gamma(v) = \gamma_0  -  \frac{dq}{c  + d v^2}.
\label{SET}
\end{equation}
Another "ansatz" is the Schienbein-Gruler formula which was empirically founded by measurements of cell motion
[29]
\begin{equation}
\gamma(v) = \gamma_0 \left( 1 - \frac{V_0}{|v|}\right).
\label{SG}
\end{equation}
The general kinetic approach to this kind of active motion has been justified in [22], where the Schienbein-Gruler formula
has been found as the particular case on the basis of microscopical kinetic theory for active friction.
A combination of the Schienbein-Gruler formula and SET-"ansatz" gives the empirical relation
\begin{equation}
\gamma(v) = \gamma_0  -  q \frac{d + a sign(v) }{c  + a v + d v^2}.
\label{SETSG}
\end{equation}
For the case $d= 0,\, c = 0,\, q=\gamma_0 V_0$ \; this reduces to the Gruler formula and for $ a=0$ to the SET formula.
Note, that for $a = 0$ Eq. (30) can be written in the in the following form [25]
\begin{equation}
\gamma(v) = \gamma_0 \frac{v^2 - \mu}{v^2 + \kappa}; \qquad \kappa \equiv \frac{c}{d}; \qquad \mu \equiv q/\gamma_0- c/d.
\label{SETDu}
\end{equation}
In the case of positive $\mu$ (note that in this Section, in contrast with Section IV, the value $\mu$ can be positive or negative; in this Section the same notations have, in general, a different sense and a different range of values. The same remark relates also to other Sections) the friction coefficient (31) equals zero at $v \equiv V_0$, where $V_0^2=\mu$.
For the most simple Gruler case, the stationary distribution reads [29]
\begin{equation}
f_s (v) =  C \exp\left[- \frac{\gamma_0}{2 D_0}\left(|v| - V_0\right)^2\right].
\label{distrf-3}
\end{equation}
Some particular results for the case $d > 0, a = 0, c=0$ have been found in [25].

The diffusion function is not so well studied. Several explicit results for dusty plasmas may be found in
[9],[10]. In following analysis we use for the diffusion coefficient $D(v)$ the simplest approximation, namely
\begin{equation}
D(v) = D_0 = const.
\end{equation}
In the special case of  a thermal heat bath we have
$ D_0 \simeq \gamma_0 T / m$, where $\gamma_0$ is the characteristic friction.
For the stationary distribution function  $f_s (v)$ in the SET-model we find for the value $a=0$, using the notations (31)
\begin{eqnarray}
f_s (v) = n Z_0^{-1} (1 + \frac{d}{c} v^2)^{\alpha} \exp[- \frac{\gamma_0}{2 D_0} v^2 ], \qquad \alpha = \frac{q}{2 D_0},
\label{distrf-4}
\end{eqnarray}
where
\begin{equation}
Z_0 = \int_{-\infty}^{+\infty} d v \exp\left(-\frac{\gamma_0}{2 D_0}[v^2 - \frac{q}{\gamma_0}  \ln(1 + (d/c) v^2)]\right).
\end{equation}
Here, as above, we use normalization of the distribution function $f_s$ of the particle density $n$.
For the stochastic potential of driven charges and the friction function (31)
we get
\begin{equation}
\phi_0(v) \simeq \frac{\gamma_0}{2 D_0}[v^2 - \frac{q}{\gamma_0} \ln(1 + (d/c) v^2)].
\label{energy}
\end{equation}
The distribution function has two different limits (see Fig. 3). In the case
$V^2_0 \gg 2 D_0/\gamma_0\equiv v^2_0$ the distribution has well expressed bistability;
in the opposite case it is more like a Maxwell distribution.
The "sum over states" $Z_0$ (35) can be rewritten, according to (34) as
\begin{equation}
Z_0 = \int_{-\infty}^{+\infty} d v (1 + \frac{d}{c} v^2)^{\alpha} \exp[- \frac{\gamma_0}{2 D_0} v^2 ].
\end{equation}
This quantity may be represented in the following convenient notation
\begin{equation}
Z_0 = v_0 \sqrt s \int_{-\infty}^\infty d\eta (1+\eta^2)^{\alpha}\exp(-s\eta^2)=v_0\sqrt{\pi s} U(\frac{1}{2}, \alpha+\frac{3}{2};s),
\label{Z0-4}
\end{equation}
where we introduced the variable $\eta\equiv v/\sqrt \kappa$ and the notations $s = \kappa/v_0^2, t = \mu/v_0^2, \alpha \equiv s+t$ and used the known representation for the hypergeometric functions ($Re a>0$;  this condition wholly covers the region of possible driven motion, where the condition $\mu>0$ is necessarily fulfilled). The function $U(a,b;z)$
\begin{eqnarray}
U(a,b;z) = \Gamma^{-1}(a) \int_0^{\infty} dt t^{a-1} (1+t)^{b-a-1} \exp(-zt); \qquad \Gamma(a) = \int_0^{\infty} dt t^{a-1} \exp(-t)\nonumber\\
U(\frac{1}{2}, \alpha+\frac{3}{2};s)=\frac{1}{\sqrt \pi} \int_0^{\infty} dt t^{-1/2} (1+t)^{\alpha} \exp(-st) \qquad \qquad
\end{eqnarray}
is sometimes called the Kummer-Tricomi function and can be represented by a linear combination of the degenerated hypergeometric function $\Phi(a,b;z)\equiv {_1}F_1(a,b;z)$.
According to [30,31], the explicit representation is, respectively
\begin{eqnarray}
U(a,b;s)\equiv \Psi(a,b;s)=\frac{\exp (s/2)}{s^{b/2}}W_{b/2 - a, (b-1)/2}(s);\qquad  \qquad \nonumber\\
U(\frac{1}{2}, \alpha+\frac{3}{2};s)=\frac{1}{\sqrt \pi} \int_0^{\infty} dt t^{-1/2} (1+t)^{\alpha} \exp(-st)=\exp (s/2)\frac{1}{s^{\frac{\alpha}{2}+\frac{3}{4}}}W_{\frac{\alpha}{2}+\frac{1}{4},\frac{\alpha}{2}+\frac{1}{4}}(s),
\end{eqnarray}
where $W_{\lambda,\mu}(s)$ is the Whittaker function.

For calculations of the averages values $<v^{2l}>$ ($l$ positive and integer) we can use the general representation related with $Z_0$
\begin{eqnarray}
<v^{2l}> =\frac{1}{Z_0}(\sqrt s v_0)^{2l+1}\int_{-\infty}^\infty d\eta \eta^{2l} (1+\eta^2)^{\alpha}\exp(-s\eta^2)=\nonumber\\ (\sqrt s v_0)^{2l}\frac{\Gamma(l+\frac{1}{2})U(l+\frac{1}{2}, l+\alpha+\frac{3}{2};s)}{\Gamma(\frac{1}{2})U(\frac{1}{2}, \alpha+\frac{3}{2};s)}.
\label{Z0-5}
\end{eqnarray}

For large values of $s$ (we use [30])
\begin{eqnarray}
U(a,b;s) \simeq \frac{1}{s^a}[1-\frac{(a_1)(1+a-b)_1}{s}],
\end{eqnarray}
where $(a)_n\equiv \Gamma(a+n)/\Gamma(a)$.
Therefore for large values of $s$ it follows from (42):
\begin{eqnarray}
U(\frac{1}{2}, \alpha+\frac{3}{2};s)=\frac{1}{\sqrt s} [1+\frac{\alpha}{2s}]; \qquad U(\frac{3}{2}, \alpha+\frac{5}{2};s)=\frac{1}{s^{3/2}} [1+\frac{3\alpha}{2s}],
\end{eqnarray}
\begin{equation}
Z_0 = \sqrt\pi v_0 [1+\frac{\alpha}{2s}].
\label{Z0-6}
\end{equation}
For small values of $s$ the representation for $U(a,b;s)$ is different for various $\alpha+\frac{3}{2}$. To find the limiting values of the "sum over states" $Z_0$ for small $s$ we consider the limits:
\begin{eqnarray}
U(\frac{1}{2}, \alpha+\frac{3}{2};s)\approx \frac{\Gamma (\alpha+1/2)}{\Gamma(1/2)} s^{-\alpha-1/2},\; Z_0\approx v_0 \Gamma (\alpha+1/2)s^{-\alpha},\; (\alpha \geq 1/2);\nonumber\\
\approx\frac{\Gamma (\alpha+1/2)}{\Gamma(1/2)} s^{-\alpha-1/2}+O(1),\, Z_0\approx v_0 \Gamma (\alpha+1/2)s^{-\alpha},\; (-1/2<\alpha<1/2);\;\nonumber\\
\approx -\frac{1}{\Gamma(1/2)}\ln[s+\psi (1/2)],\;Z_0\approx -v_0 s^{1/2}\ln[s+\psi (1/2)],\;(\alpha=-1/2);\nonumber\\
\approx\frac{\Gamma (-\alpha-1/2)}{\Gamma(-\alpha)}+O(s^{-\alpha-1/2}),\; Z_0 \approx v_0 \frac{\Gamma (-\alpha-1/2)}{\Gamma(-\alpha)}\sqrt{\pi s},\;(-3/2<\alpha<-1/2);\;\nonumber\\
\approx\frac{1}{\Gamma(3/2)}+O(s \ln s),\; Z_0\approx v_0 \frac{1}{\Gamma(3/2)}\sqrt{\pi s},\;(\alpha=-3/2);\;\nonumber\\
\approx\frac{\Gamma (-\alpha-1/2)}{\Gamma(-\alpha)}+O(s),\;Z_0 \approx v_0 \frac{\Gamma (-\alpha-1/2)}{\Gamma(-\alpha)}\sqrt{\pi s},\; (\alpha <-3/2),\;
\end{eqnarray}

As we already mentioned above, the distribution function (34) for $\mu<0$ (passive friction) has quasi-Maxwellian form with a maximum at $v=0$; for the case of driven motion $\mu>0$ the distribution function is bistable with a maximums at $v=\pm\sqrt{\mu }$ (Fig.3). For the case of driven motion $\alpha=q/2D_0=\gamma_0(\kappa+\mu)/2D_0>0$, since $\mu>0$

\begin{figure}[htbp]
\begin{center}
\includegraphics[width=4cm,height=4cm]{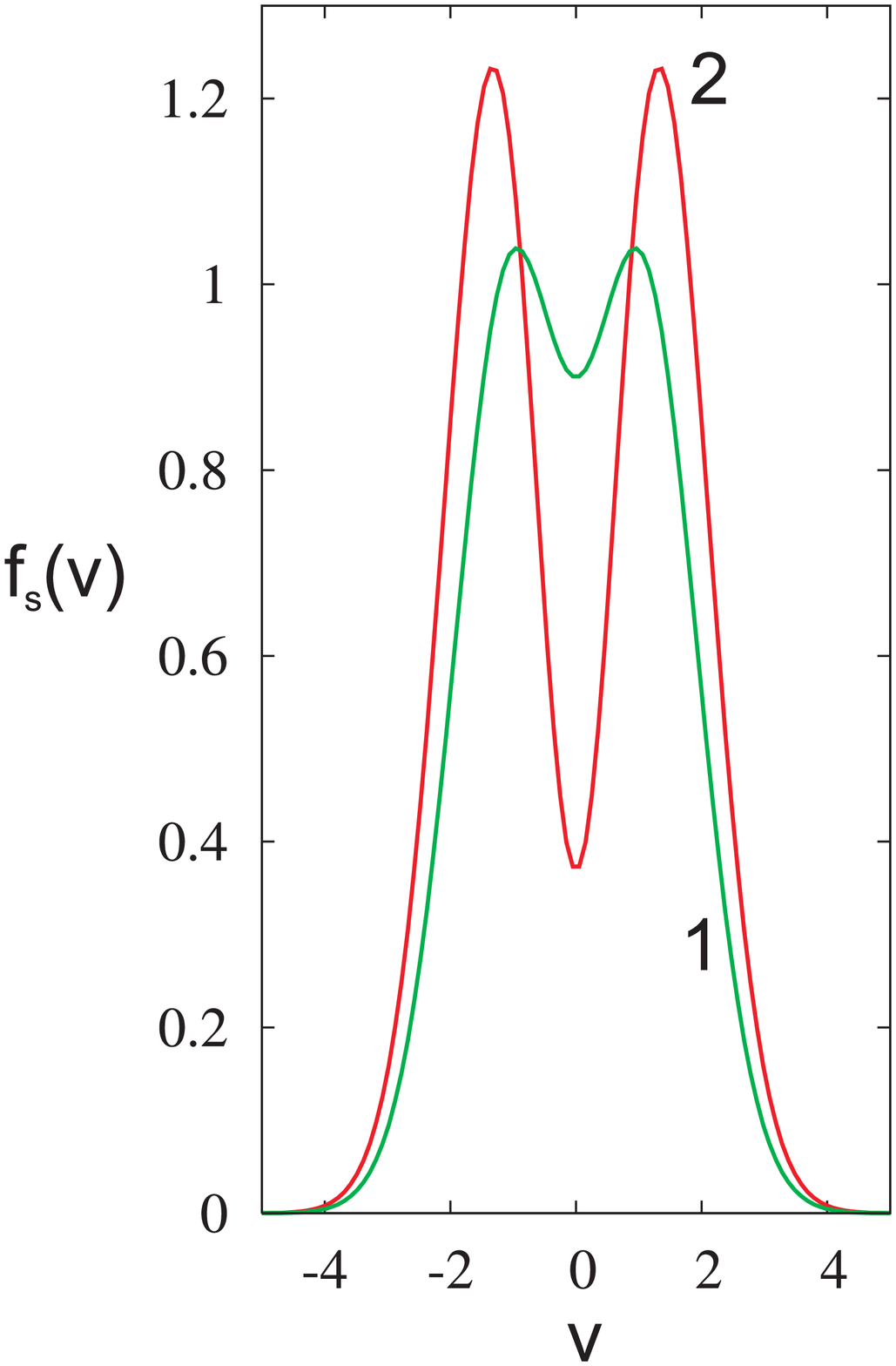}
\includegraphics[width=4cm,height=4cm]{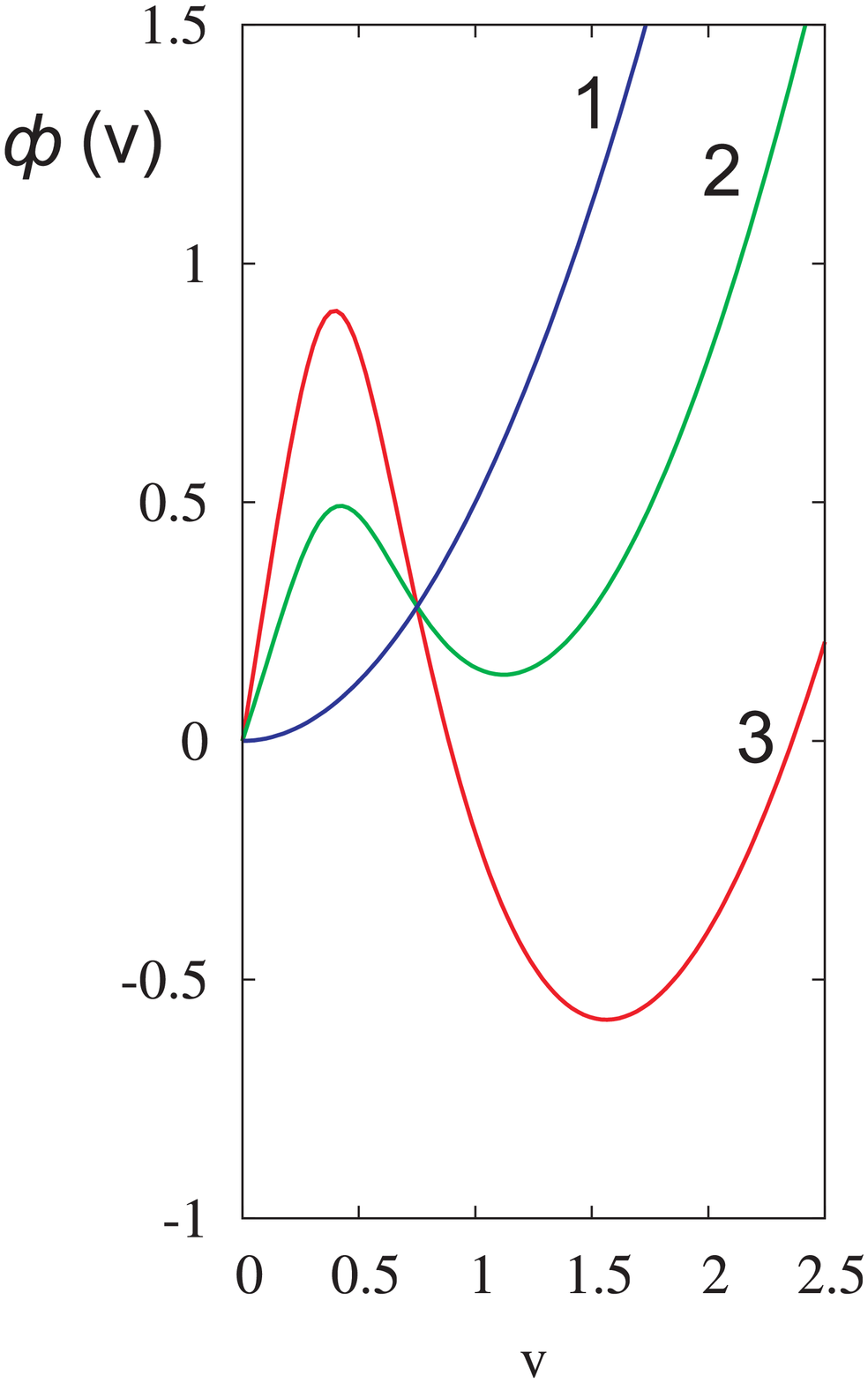}
\end{center}
\caption{The left panel shows a graph of two typical Gruler-type bistable velocity distributions: the curves 1 and 2 represent the distribution function $f_s(v)$ according to Eq.(32) for $V_0^{(1)}<V_0^{(2)}$.
The right panel shows a graph of the corresponding shapes of the stochastic potential of driven particles defined by Eq. (27). Curve 1 corresponds to the Maxwell distribution, curves 2 and 3 correspond to the Gruler-type distribution with $V_0^{(2)}<V_0^{(3)}$ respectively.
}
\label{veldis}
\end{figure}


Similarly to the Gibbs theory, the quantity $Z_0$ ("partition function") is a useful tool for many applications.
For example the dispersion of the distribution may be expressed as a derivative ($\kappa \equiv s v_0^2$) of $Z$ with respect of $s$
\begin{equation}
\langle v^2 \rangle = -s v_0^2 \left(\frac{\partial \ln Z_0 (\alpha,s)}{\partial s}\right)_{ \alpha\equiv s+t=const.}=\frac{s v_0^2}{2} \frac{ U(\frac{3}{2}, \alpha+\frac{5}{2};s)}{U(\frac{1}{2}, \alpha+\frac{3}{2};s)}.
\end{equation}
For large values of $s$ the universal asymptotic behavior $Z_0$ is given by (44) and we then find
 \begin{equation}
\langle v^2 \rangle = \frac{v_0^2}{2} \frac {1+3\alpha/2s}{1+\alpha/2s}.
\end{equation}

Now we consider simple estimation of the quantity $\langle v^2\rangle$ for a small $s$. For active motion $\mu>0$ the value $\alpha$ is always positive and, therefore, to calculate $\langle v^2 \rangle$ approximately in this case we may use the two first lines in (45) to find
\begin{equation}
\langle v^2 \rangle =\alpha v_0^2.
\end{equation}
This representation is also correct for passive friction ($\mu<0$) if $\alpha=s+t>0$.

Since for $\alpha=0$ and for arbitrary values of $s$ the functions  $U(\frac{1}{2}, \alpha+\frac{3}{2};s)=U(\frac{1}{2},\frac{3}{2};s)=1/\sqrt s$ and $U(\frac{3}{2}, \frac{5}{2};s)=U(\frac{1}{2},\frac{3}{2};s)=1/s^{3/2}$ we arrive at $Z_0=v_0 \sqrt \pi$. This case describes the equilibrium stationary solution (7). It is easily verified that for equilibrium the explicit value of $\langle v^2\rangle$ equals
\begin{equation}
\langle v^2 \rangle =\frac{ v_0^2}{2}.
\end{equation}
All particular cases of various $\alpha$-values and small values of $s$ can be considered by using (45).

\subsection{Nonlinear drift of driven charges including external fields}

In the presence of electrical fields the stationary
Fokker-Planck equation is in our approximation given by Eq. (4) and the general stationary solution is given by Eq. (8).

The typical behavior of the distribution function $f_s(v)$ is shown graphically in Fig. 4 for various values of the homogeneous electric field.

\begin{figure}[htbp]
\begin{center}
\includegraphics[width=6cm,height=6cm]{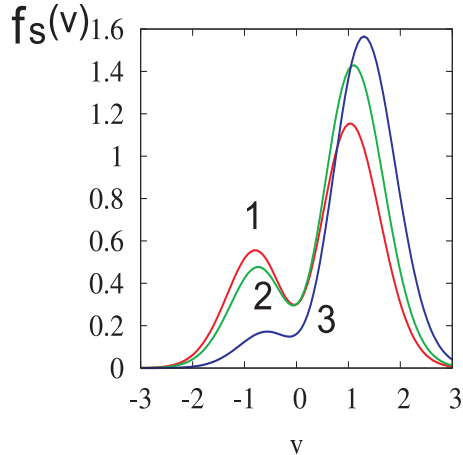}
\end{center}
\caption{Typical velocity distributions for three different electrical fields ($\varepsilon_1< \varepsilon_2<\varepsilon_3$) according to Eq.(8). The graph shows schematically the shape of the distributions for increasing fields as a function
of the velocity $v$ for $\varepsilon_1$ (curve 1),  $\varepsilon_2$ (curve 2), and  $\varepsilon_3$ (curve 3). It is observed that the distribution is shifted by an external field to the field direction. With increasing field strength $\varepsilon$ the left maximum
disappears and the right maximum increases. In strong fields the distribution possesses only one maximum.}
\label{velodisfields}
\end{figure}


The calculation of the nonlinear drift velocity for driven particles
is in general a complicated problem.
An analytical solution was given so far only for the model studied by Dunkel et al.
[25]. This model corresponds to a SET model with $\kappa=0$ (or $c/d\rightarrow 0$).
The solution is expressed by hypergeometric functions [25].
In general there are two limits which can be easy calculated:
(i) $E \rightarrow 0$ and (ii) $E \rightarrow \infty$.\\
Let us first consider the low field limit, which is the range of Ohm's law, where the current is proportional to the
field $j = \sigma E$. By linearization of the distribution function
Eq.(8) and assuming $D(v) = D_0$ we obtain for the mean velocity
\begin{eqnarray}
\langle v \rangle = \int dv v f_s (v, E) = \int dv v f_s (v, E=0) \left[1 - v \frac{e E }{m D_0}+O(E^2)\right].
\label{(vE)}
\end{eqnarray}
In the linear approximation one derives
\begin{equation}
\langle v \rangle_{linear\, on\, E} =   \frac{|e| E }{m D_0} \langle v^2 \rangle_{E=0}.
\end{equation}
This result, which expresses a special form of the fluctuation-dissipation theorem
reduces the calculation of the Ohmic current to the dispersion of the distribution for the case of zero field.

For the friction function in the form (31) we find - by using linear approximation on the electrical field, see (51) - for the conductivity in terms of the functions $U(a,b;s)$ the following:
\begin{equation}
\sigma = \frac{n e^2}{m D_0} \langle v^2 \rangle_{E=0} =\frac{n e^2}{m D_0}\frac{s v_0^2}{2} \frac{ U(\frac{3}{2}, \alpha+\frac{5}{2};s)}{U(\frac{1}{2}, \alpha+\frac{3}{2};s)}.
\end{equation}

In the approximation of maximal driven velocity ($\alpha \gg 1$) explained above we find (see (48)) for the Ohmic conductivity
\begin{equation}
\sigma \simeq \frac{n e^2}{m D_0}\alpha v_0^2.
\end{equation}
In relation to the Drude conductivity this gives the estimate
\begin{equation}
\frac{\sigma}{\sigma_D} \simeq 2\alpha.
\end{equation}
Note, that the maximal velocity changes in a broad interval between the thermal velocity and driven velocity $V_0$ as follows form Fig. 4 and Eq. (34).
Therefore, we show that driving effects may lead to a strong enhancement of
the Ohmic conductivity for $\alpha >1$.

Let us consider qualitatively the opposite case of large fields. Then the distributions shown in Fig. 4
are getting mono-stable since $\gamma(v) \simeq \gamma_0$ at $v \gg v_0$. In this limit the distribution approaches the form equivalent to (14)
\begin{eqnarray}
f_s (v,E) = n Z_0^{-1} \exp\left[ - \frac{\gamma_0}{2 D_0} (v - v_D)^2 \right], \qquad v_D = \frac{|e| E}{m \gamma_0}.
\label{distrEbig}
\end{eqnarray}
Here $v_D$ is the Drude velocity. This gives for the mean velocity
\begin{equation}
\langle v \rangle =   \frac{|e| E }{m \gamma_0} = v_D,
\end{equation}
and the conductivity is similar to the Drude conductivity (15).
This result proves that in the limit of large fields the system behaves like a Drude system,
the influence of driven motion of charges is negligible. Knowing the behavior of conductivity at low and at high electric fields, it remains to study the intermediate region.

In the following we use introduce the electrical field $E$ via the variable $\epsilon$ which is equivalent to the Drude velocity
$\epsilon = \mid e\mid E / m \gamma_0\equiv v_D$ or via the dimensionless variable $\varepsilon=e E \sqrt \kappa /m D_0$.
This leads in the Drude conductivity (see introduction) for linear approximation on field $E$.

Typical curves for the drift velocity of nonlinear systems in an electrical field
show changing slopes [28]. An observation often described is that a
steep linear increase for small field values $\varepsilon$ is followed by a plateau which is then followed again by a linear
increase at higher fields $\varepsilon$ [6],[13],[15]-[17].
Most of these observation refer to polaron-type systems and they are based on numerical simulations
and analytical estimates. A full theoretical explanations of these findings is still lacking.

We will give now a systematic derivation of the drift and the conductivity in the electrical field
for models of nonlinear on electric field friction function.
In the following we concentrate on the SET-model which seems to be most realistic.

We will do several transformations and use the following non-dimensional variables
$\eta \equiv v / \sqrt \kappa$, $s \equiv \kappa / v_0^2$, $t=\mu/v_0^2$, $\alpha=s+t$, $\varepsilon=e E \sqrt \kappa /m D_0= 2 v_D \sqrt \kappa/v_0^2\equiv 2\epsilon \sqrt \kappa/v_0^2 \equiv 2\epsilon \sqrt s/v_0$. The variables $\varepsilon$ and $\epsilon$ are proportional to the electric field.

We use the general expression for the nonlinear current in the considered model:
\begin{eqnarray}
j=
-e \frac{n v_0 \sqrt s \int_{-\infty}^\infty d\eta \eta \left(1+\eta^2\right)^{\alpha}\exp \left(-s \eta^2- \varepsilon \eta\right)}{\int_{-\infty}^\infty  d \eta(1+\eta^2)^{\alpha}\exp (-s\eta^2-\varepsilon \eta)}.
\label{A59}
\end{eqnarray}

Since the nominator is the derivative of the denominator, a more convenient form may be obtained by using the procedure similar to the Gibbs method in thermodynamics, by writing
\begin{eqnarray}
j=
e n v_0 \sqrt s \frac{\partial \ln Z(\varepsilon, \alpha,s)}{\partial \varepsilon},
\label{A60}
\end{eqnarray}
where the "sum over states" is
\begin{eqnarray}
Z (\varepsilon, \alpha, s)
= \int_{-\infty}^\infty  d \eta (1+\eta^2)^{\alpha}\exp (-s \eta^2- \varepsilon \eta).
\label{sumoverstates-6}
\end{eqnarray}
The whole problem is now reduced to estimating one function $Z$ depending on the variable $\varepsilon$ and on two parameters
$\alpha$ and $s$ and the partial derivatives of $Z$.

Let us now consider the case of integer $\alpha$-values in more detail. For the case $\alpha=1$ we find the explicit result for arbitrary value of $\varepsilon$
\begin{eqnarray}
Z (\varepsilon, 1, s)
= \int_{-\infty}^\infty  d \eta (1+\eta^2)\exp (-s \eta^2- \varepsilon \eta).
\label{sumoverstates-7}
\end{eqnarray}
The explicit result for $Z(\varepsilon, 1, s)$ in this case reads
\begin{eqnarray}
Z (\varepsilon, 1, s)
= \int_{-\infty}^\infty  d \eta (1+\eta^2)\exp (-s \eta^2- \varepsilon \eta)=\exp (\varepsilon^2/4s)\sqrt{\pi/s}[1+(\varepsilon^2/4s^2+1/2s)].
\label{sumoverstates-8}
\end{eqnarray}
The corresponding nonlinear current for an arbitrary electric field (or arbitrary $\varepsilon$) can be easily calculated
\begin{eqnarray}
j(\varepsilon,1,s)=
e n v_D [1+\frac{1}{s(1+\varepsilon^2/4s^2+1/2s)}].
\label{A61}
\end{eqnarray}
For both limiting cases $s\rightarrow 0$ and $s\rightarrow\infty$ the nonlinear current $j(\varepsilon,1,s)$ tends to the Drude result $j_D=e n v_D$.

For the case of a weak electric field (allowing for linear approximation), equation (62) gives for the linearized current $j_L(\varepsilon,1,s)$:
\begin{eqnarray}
j_L(\varepsilon,1,s)=
e n v_D [1+\frac{2}{1+2s}].
\label{A62}
\end{eqnarray}
Therefore, in the linear approximation there is an increase of the linear conductivity:
\begin{eqnarray}
\frac{j_L(\varepsilon,1,s)}{j_D}\equiv\frac{\sigma(1,s)}{\sigma_D}=[1+\frac{2}{1+2s}]\equiv [1+\frac{2 \tau}{1+\tau}].
\label{A63}
\end{eqnarray}
For the cases of active ($\mu>0$) and passive ($\mu<0$) friction in the limit $s\gg 1$ the conductivity tends to the Drude result $\sigma(1,s\gg 1)\rightarrow \sigma_D$. For a small values $s\ll1$ the linear conductivity tends to the maximum $\sigma(1,s\ll 1)\rightarrow 3 \sigma_D$.

For a large electric field $\varepsilon\gg 1$, equation (62) also implies a linear behavior of the current similar to the Drude result:
\begin{eqnarray}
\frac{j(\varepsilon\gg 1,1,s)}{j_D}\equiv\frac{\sigma(1,s)}{\sigma_D}=1.
\label{A64}
\end{eqnarray}

Let us now consider the case $\alpha=2$. For $Z (\varepsilon, 2, s)$ we arrive at the expression
\begin{eqnarray}
Z (\varepsilon, 2, s)=
\exp (\varepsilon^2/4s)\sqrt{\pi/s}[1+(\varepsilon^2/2s^2+1/s)+(\varepsilon^4/16 s^4+3 \varepsilon^2/4 s^3+3 /4 s^2)].
\label{sumoverstates-9}
\end{eqnarray}

The corresponding nonlinear current for $\alpha=2$ is given by
\begin{eqnarray}
j(\varepsilon,2,s)=
e n v_D \times \frac{ [1+(\varepsilon^2/2s^2+1/s)+
(\varepsilon^4/16 s^4+3 \varepsilon^2/4 s^3+3 /4 s^2)+\frac{2}{s}+\frac{\varepsilon^2}{2 s^3}+\frac{3}{ s^2}]}{[1+(\varepsilon^2/2s^2+1/s)+(\varepsilon^4/16 s^4+3 \varepsilon^2/4 s^3+3 /4 s^2)]}.
\label{A65}
\end{eqnarray}
In linear approximation this becomes
\begin{eqnarray}
\frac{j_L(\varepsilon,2,s)}{j_D}\equiv\frac{\sigma(2,s)}{\sigma_D}=
1+\frac{12+8s}{3+4s+4s^2}.
\label{A66}
\end{eqnarray}
For the case $\alpha=2$, as for the case $\alpha=1$, the nonlinear conductivity $\sigma(2,s)$ tends to $\sigma_D$ for both limiting values of the parameter $s$, namely for $s\rightarrow\infty$ and $s\rightarrow 0$. For the case of linear conductivity the corresponding current $j_L(\varepsilon,2,s\rightarrow\infty)\rightarrow 0$ and $j_L(\varepsilon,2,s\rightarrow 0)\rightarrow 5$. Therefore, the maximum linear conductivity exists at the value $s=0$ and equals $\sigma(2,0)=5 \sigma_D$.

For a large electric field $\varepsilon\gg 1$ equation (68) also implies a linear behavior of the current $j(\varepsilon \gg 1,2,s)\rightarrow j_D$, similar to the Drude result, as for the case $\alpha=1$.

It is also interesting to calculate the linear current for an arbitrary value of the parameter $\alpha$. In this case we need only the functions
\begin{eqnarray}
Z (\varepsilon=0, \alpha, s)
= \int_{-\infty}^\infty  d \eta (1+\eta^2)^{\alpha}\exp (-s \eta^2)
\label{sumoverstates-10}
\end{eqnarray}
and
\begin{eqnarray}
\partial Z (\varepsilon, \alpha, s)/\partial \varepsilon \simeq
 \varepsilon \int_{-\infty}^\infty  d \eta \eta^2 (1+\eta^2)^{\alpha}\exp (-s \eta^2)+ O(\varepsilon)=\nonumber\\
 \varepsilon [Z (\varepsilon=0, \alpha+1, s)-Z (\varepsilon=0, \alpha, s)]+ O(\varepsilon).
\label{sumoverstates-11}
\end{eqnarray}
The current in the case under consideration equals
\begin{eqnarray}
j(\varepsilon,\alpha,s)=\frac{e n v_0 \sqrt s}{Z(\varepsilon=0,\alpha,s)}\frac{\partial Z(\varepsilon,\alpha,s)}{\partial \varepsilon}=
=e n v_D s\, \frac{ U(3/2,\alpha+5/2;s)}{U(1/2, \alpha+3/2;s)}.
 \label{A67}
\end{eqnarray}
It is easy to see that this result is identical to Eq. (52). Some particular cases of simplification for the functions $U(m/2, \alpha,s)$ for various $\alpha$-values are presented in [31].

Taking into account the general relation between the functions $U(a,b;z)$ and $W_{k,\mu(z)}$ [31]
\begin{eqnarray}
W_{k,\mu(z)}=\exp(-z/2)z^{1/2+\mu}U(1/2+\mu-k, 1+2\mu;z), \; k=b/2-a, \qquad \mu=b/2-1/2.
 \label{A68}
\end{eqnarray}
and the particular relations $W_{\alpha/2-1/4,\alpha/2+3/4}(s)=\exp(-z/2)s^{\alpha/2+5/4}U (3/2,\alpha+5/2;s)$ and $W_{\alpha/2+1/4, \alpha/2+1/4}(s)=\exp(-z/2)s^{3/4+\alpha/2}U (1/2,\alpha+3/2;s)$ we obtain
\begin{eqnarray}
j(\varepsilon\rightarrow 0,\alpha,s)=e n v_D s \frac{U (3/2,\alpha+5/2;s)}{U (1/2,\alpha+3/2;s)}=e n v_D \sqrt s \frac{W_{\alpha/2-1/4,\alpha/2+3/4}(s)}{W_{\alpha/2+1/4, \alpha/2+1/4}(s)}.
 \label{A70}
\end{eqnarray}

\section{Conclusions}

The shape of nonlinear conductivity curves may be quite complicated, depending on the particular sets of the parameters which are responsible for the type of the velocity distribution functions. Figure 5 gives a graphical representation of the dependence of the conductivity on the parameters $s$ and $\varepsilon$ for $\alpha$ = 1 and $\alpha$= 2, according to the analytical results given above. The left panel of Fig. 5 shows the low field conductivity in relation to the Drude
conductivity for the two values of $\alpha = 1,2$ as a function of the parameter $1/2s$. This parameter
may be considered as a kind of "effective temperature" since it determines the dispersion
of the distribution. We observe a strong increase of the low field conductivity in relation
to the Drude value for larger values of $1/2s$. This means physically that nonlinear effects
may increase low field conductivity substantially. The right panel of Figure 5 shows the dependence
on the dimensionless field $\varepsilon$. The highest value of the derivative (differential conductivity)
is observed for low fields, then the derivative decreases monotonically with the field from a
highest value (zero field conductivity) to the Drude value of conductivity (equal to unity in
the notation in the right panel of Fig. 5, with $\sigma_D=1$ and $j_D=\varepsilon$).

\begin{figure}
\begin{center}
\includegraphics[width=5cm,height=5cm]{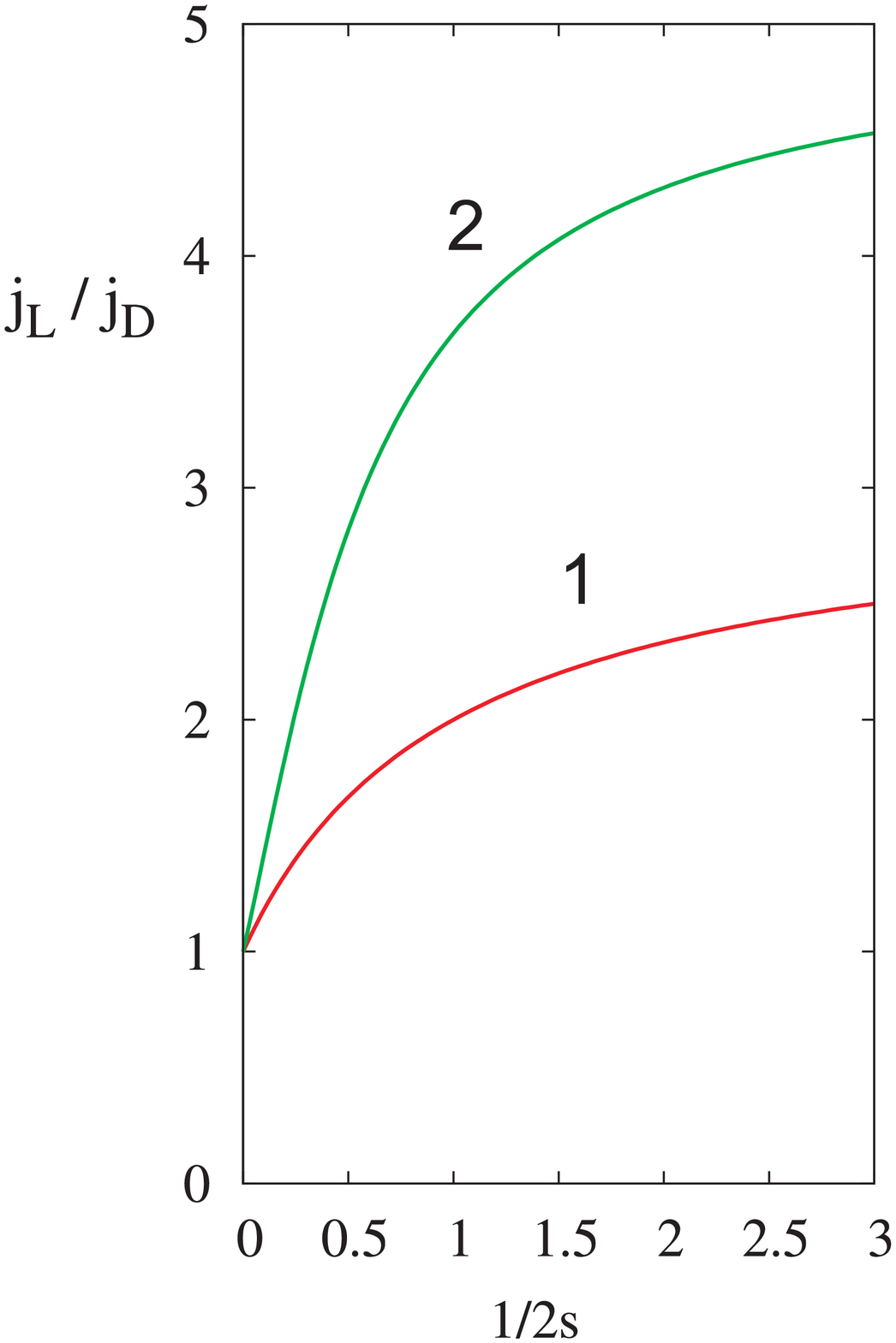}
\includegraphics[width=5cm,height=5cm]{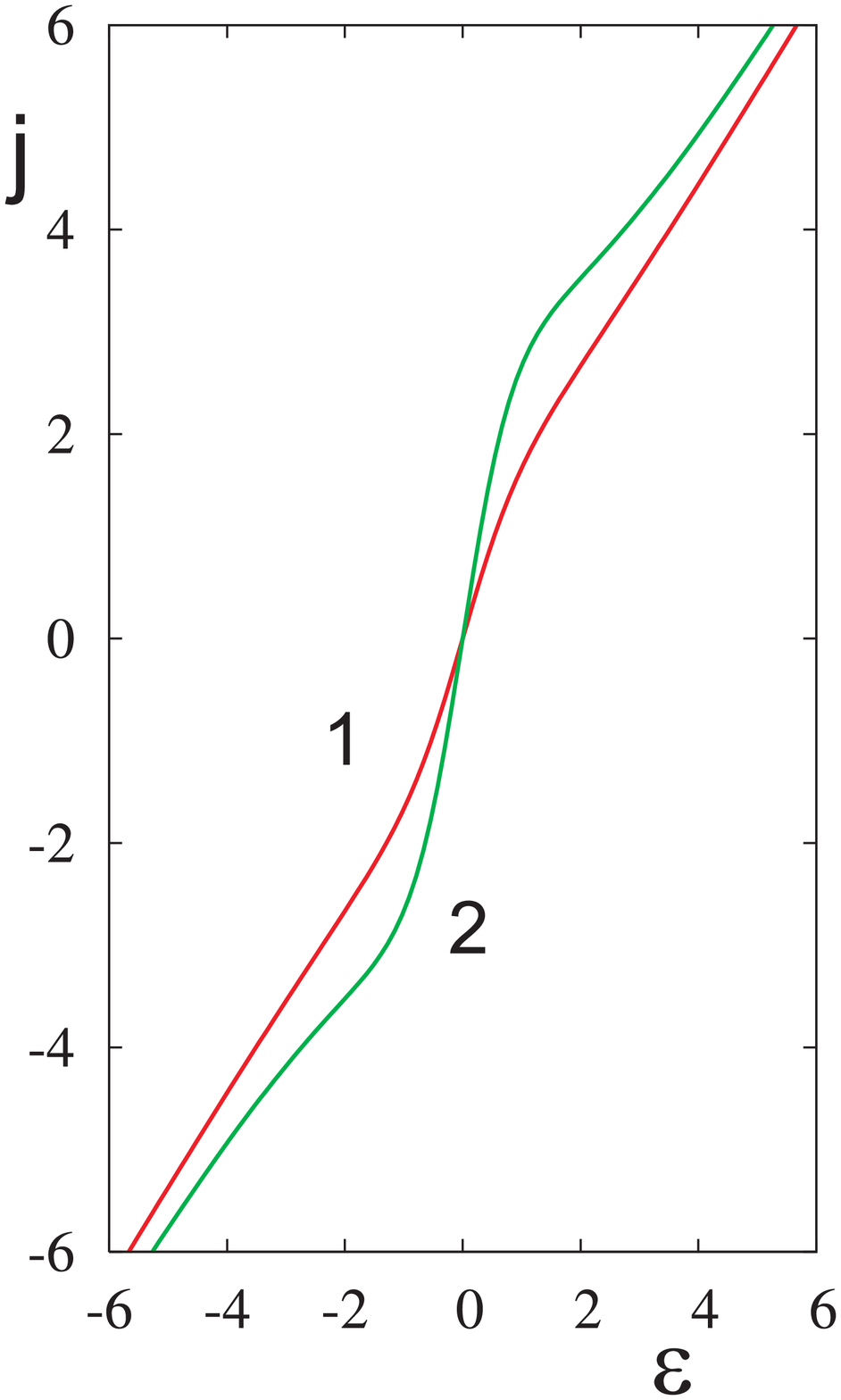}
\end{center}
\caption{Graphs showing the typical relation of the conductivity to the Drude conductivity.
Left panel: Monotonic increase of the ratio of the linear current (field independent conductivity) to the Drude current with decreasing parameter $s$ (curves 1 and 2 represent alpha = 1 and 2, respectively). Right panel: Increase of the current with
increasing dimensionless field $\varepsilon$. The curves correspond to
 $\alpha = 1, s = 1/2$ (curve 1) and $\alpha = 2, s=1/2$ (curve 2) for $j_D=\varepsilon$.}
\label{Fig.5}
\end{figure}

\begin{figure}
\begin{center}
\includegraphics[width=\textwidth]{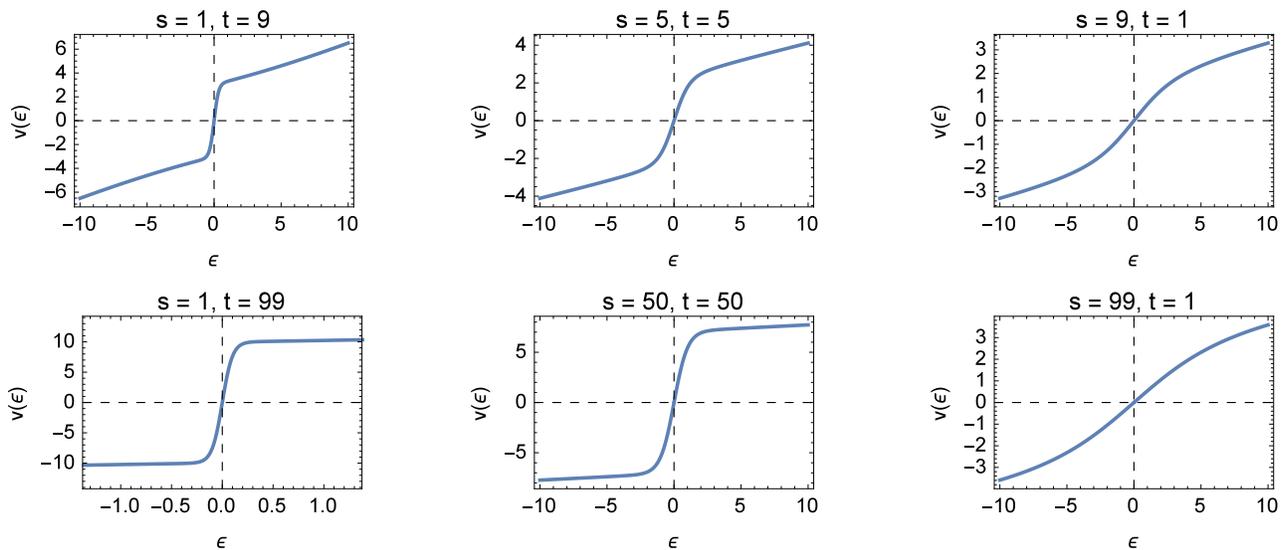}
\end{center}
\caption{Non-Ohmic dimensionless velocity obtained from numerical evaluations versus the dimensionless electrical field
for several combinations of the parameters $s$ and $t$. Note that here $\alpha = s+t$.
}
\label{Fig.6}
\end{figure}

The result of numerical evaluations of the current for a large set of values of the parameters $\alpha$ and $s$
is shown graphically in Fig. 6. The panels show the behavior of $v(\varepsilon)$ for various values of the parameters $s$ and $t = \alpha - s$.  The graphs reveal that the typical shape of the $v(\varepsilon)$ profile is similar to that in the cases $\alpha$ = 1 and $\alpha$=2. However, they also demonstrate some new aspects, such as the individual dependence on the parameters $t$
and $s$, which gives rise to different shapes of the curves. This property demonstrates
that the characteristic dependence on the field may be very different and depends sensitively on the particular set of $t$ and $s$-values. This sensitivity provides the interesting possibility to shape a "characteristics on demand", just by choosing particular values of $t$ and $s$. The problem of providing a "current-voltage characteristics on demand" might be of interest for applications
to problems of nonlinear electronics. The generalization of the Fokker-Planck theory developed in this paper may provide the tools for constructing a particular desired characteristics.

By our analysis, based on the Fokker-Planck models, we have shown in the present paper that nonlinear
effects may substantially increase low field conductivity. Typically, as shown by the examples
given in Figs. 5 and 6, the increase reduces at high electrical fields again to the value provided
by the linear Drude theory. Therefore, the search for highly conducting
materials should include nonlinear effects and the present generalized Fokker-Planck model
may give important hints for search of highly conducting materials by variations of the
material and system parameters (such as the  effective temperature D and the friction $\gamma(v)$).
Furthermore, we have shown that the nonlinear theory may give important tools how to shape a
"current-voltage characteristics on demand".

\section*{Acknowledgment}
The authors are thankful to Alexander P. Chetverikov,
Lutz Schimansky-Geier, Pieter Schram, Igor Sokolov, Manuel G. Velarde for valuable discussions

Sergey Trigger is thankful for support to Russian Science Foundation
(project no. 14-19-01492).


\begin{thebibliography}{99}




[1] G. R\"opke, Nonequilibrium Statistical Physics, Wiley, 2013\\


[2] H. Falkenhagen, Theorie der Elektrolyte, Hirzel Verlag, Leipzig 1971\\


[3] V.P. Silin, \emph{Dokl. Akad. Nauk SSSR, Fizika} \textbf{161}, No 2, 328-331 (1965)\\



[4] Yu.L. Klimontovich, Kinetic Theory of Nonideal Gases and Nonideal Plasmas, Pergamon Press, Oxford, London-New-York, 1982\\
%

[5] K.J. Donovan, E.G. Wilson, \emph{Phil. Mag. B} \textbf{44}, 31-45 (1981)\\
%

[6] A.A. Gogolin, \emph{Physics Reports} \textbf{157}, 348--391 (1988)\\
%

[7] V. Fortov, I. Yakubov and A. Khrapak, Physics of
Strongly Coupled Plasma (Clarendon Press, Oxford) 2006\\


[8] A.G.Zagorodny, P.P.J.M.Schram, S.A.Trigger, \emph{Phys.Rev.Lett.} \textbf{84} p. 3594 (2000)\\


[9] S.A. Trigger, W. Ebeling, A.M. Ignatov I.M. Tkachenko, Contr. Pl. Phys., N5-6, 377 (2003)\\


[10] J. Dunkel, W. Ebeling, S.A. Trigger, \emph{Phys. Rev E} \textbf{70}, 046406 (2004)\\


[11] V.A. Makarov, E. Del Rio, W. Ebeling, and M.G. Velarde, \emph{Physical Review E} \textbf{64},
0366601-36615 (2001)\\
%

[12] E. Del Rio, V.A. Makarov, M.G. Velarde, and W. Ebeling, \emph{Physical Review E} \textbf{67}, 056208-056217 (2003)\\
%

[13] V.A. Makarov, M.G. Velarde, A.P. Chetverikov, W. Ebeling, \emph{Phys. Rev. E}  \textbf{73}, 066626-1-12 (2006)\\
%

[14] D. Hennig, A. Neissner, M.G. Velarde, W. Ebeling,
\emph{Phys. Rev. E} \textbf{73}, 024306 (2006)\\
%

[15] D. Hennig, A.P. Chetverikov,  M.G. Velarde, W. Ebeling, \emph{Phys. Rev. E} \textbf{76}, 046602 (2007)\\


[16] A.A. Gogolin,  \emph{Pis'ma Zh. Exp. Teor. Fiz.} \textbf{43}, 395 (1986)\\



[17] V.D. Lakhno, \emph{Int. J. Quant. Chem. } \textbf{110}, 127--137 (2010)\\

[18] A.P. Chetverikov, W. Ebeling, M.G. Velarde, \emph{Eur. Phys. J. B} \textbf{80}, 137 - 145 (2011)\\


[19] A.P. Chetverikov, W. Ebeling, M.G. Velarde, \emph{Eur. Phys. J. B} (2012), DOI: 10.1140/epjb/e2012-30276-x\\


[20] S.I. Braginskii, Problems of plasma theory, Atomizadat, Moscow 1963\\


[21] S.A. Trigger, \emph{Phys. Letters A} \textbf{374}, 134 (2009); ArXiv 0907.2793 v1, [cond-matt. stat.-mech.], 16
July 2009\\


[22] S.A. Trigger, \emph{Phys. Rev.} \textbf{E67}, 046403 (2003).\\


[23] S.A. Trigger, W. Ebeling, G.J.F. van Heijst, P.P.J.M. Schram, I.M. Sokolov, \emph{Physics of Plasmas }\textbf{17}, 042102 (2010)\\


[24] A.A. Dubinova, S.A. Trigger, \emph{Physics Letters} \textbf{A 376}  1930 (2012)\\



[25] J. Dunkel, W. Ebeling, U. Erdmann, \emph{Eur. Phys. J. B} \textbf{24}, 511-524 (2001)\\


[26] W. Ebeling, I. Sokolov, Statistical thermodynamics and stochastic theory of nonequilibrium systems, Singapore 2005\\


[27] B. Dybiec, E. Gudowska-Nowak, I.M. Sokolov, \emph{Phys. Rev. E} \textbf{78}, 011117 (2008)\\


[28] W. Ebeling, A. Chetverikov, M.G. Velarde, Proc. Int. Conf. ICENET2012, MIPT Dolgoprudnyi 2012 (www.icenet2012.net)\\

[29] M. Schienbein, K. Franke, and H. Gruler, \emph{Phys.Rev E}
\textbf{49}, 5462 (1994).\\


[30] I.S. Gradshtein and I. M. Rijik, Tables of Integrals, Sums, Series and Products,
Nauka, Moscow, 1971.\\


[31] Handbook of Mathematical Functions, Edited M.Abramowitz and I.A.Stegun, National Bureau Standards, 1964.\\

\end{thebibliography}
\end{document}